\documentclass{article}

\usepackage[toc,page]{appendix}
\usepackage[a4paper, margin=1.1in]{geometry}
\usepackage{graphicx}%
\usepackage{dcolumn}%
\usepackage{bm}%
\usepackage[version=3]{mhchem} %
\usepackage{amsmath}

\usepackage{xr}

\usepackage[utf8]{inputenc}
\usepackage[T1]{fontenc}
\usepackage{mathptmx}
\usepackage{SIunits}
\usepackage{xcolor}
\usepackage{subcaption}

\usepackage{array}
\newcolumntype{P}[1]{>{\centering\arraybackslash}p{#1}}
\usepackage{fancyhdr}

\usepackage{float}
\usepackage{multirow}
\usepackage{rotating}
\usepackage{adjustbox}
\usepackage{makecell}
\usepackage{caption}

\usepackage{changepage}

\usepackage{amsmath}
\usepackage{url}
\usepackage[labelsep=space,labelfont=bf]{caption}

\title{Molecular simulations to investigate the impact of N6-methylation in RNA recognition: Improving accuracy and precision of binding free energy prediction}
\author{Valerio Piomponi$^{1,2}$ \and Miroslav Krepl$^3$ \and Jiri Sponer$^3$ \and Giovanni Bussi$^1$}

\date{%
    $^1$Scuola Internazionale Superiore di Studi Avanzati, SISSA, via Bonomea 265, 34136 Trieste, Italy\\%
    $^2$Area Science Park, località Padriciano, 99, 34149 Trieste, Italy\\%
    $^3$Institute of Biophysics of the Czech Academy of Sciences, Kralovopolsk\'a 135, 612 00 Brno, Czech Republic\\%
}

\begin{document}

\maketitle

\begin{abstract}
N6-methyladenosine (m$^6$A) is a prevalent RNA post-transcriptional modification that plays crucial roles in RNA stability, structural dynamics, and interactions with proteins. The YT521-B (YTH) family of proteins, which are notable m$^6$A readers, function through their highly conserved YTH domain. Recent structural investigations and molecular dynamics (MD) simulations have shed light on the recognition mechanism of m$^6$A by the YTHDC1 protein.
Despite advancements, using MD to predict the stabilization induced by m$^6$A on the free energy of binding between RNA and YTH proteins remains challenging, due to inaccuracy of the employed force field and limited sampling. For instance, simulations often fail to sufficiently capture the hydration dynamics of the binding pocket.
This study addresses these challenges through an innovative methodology that integrates metadynamics, alchemical simulations, and force-field refinement. Importantly, our research identifies hydration of the binding pocket as giving only a minor contribution to the binding free energy and emphasizes the critical importance of precisely tuning force-field parameters to experimental data. By employing a fitting strategy built on alchemical calculations, we refine the m$^6$A partial charges parameters, thereby enabling the simultaneous reproduction of N6 methylation on both the protein binding free energy and the thermodynamic stability of nine RNA duplexes.
Our findings underscore the sensitivity of binding free energies to partial charges, highlighting the necessity for thorough parameterization and validation against experimental observations across a range of structural contexts.
\end{abstract}

\section{Introduction}

N6-methyladenosine (m$^6$A) is the most common post-transcriptional modification found in nature, and is widespread in both coding and noncoding RNAs \cite{gilbert2016messenger,harcourt2017chemical,patil2016m,he2021m6a}. The methylation of the N6 amino group of adenosine can affect RNA stability and structural dynamics,
as well as regulate RNA interactions with proteins.
Among those, the YT521-B (YTH) family of proteins (acting as m$^6$A readers) stands out as the most prominent and extensively examined  \cite{wang2014n6methyladenosine, zhang2010yth, luo2014molecular}. It recognizes m$^6$A via a highly conserved YTH domain \cite{wang2014n6methyladenosine, wang2015n6methyladenosine, meyer2012comprehensive, wu2017readers, reichel2019marking, zaccara2019reading, zhang2010yth}.
Its role in m$^6$A recognition has been extensively examined in recent years for the YTH domain of the YTHDC1 protein, for which several structures have been solved and deposited in the protein databank, with different oligonucleotides bound \cite{theler2014solution, li2021atomistic, li2019flexible}. All these structures show m$^6$A being recognized by a deep aromatic cage formed by protein residues, with the flanking nucleotides bound on the protein surface. Several studies have used molecular dynamics (MD) simulations to investigate how the protein binds and recognizes m$^6$A \cite{krepl2021recognition, li2019flexible, li2021atomistic, li2022structure, zhou2022specific}
(see \cite{piomponi2022MD} for a recent review). All these works visualize how the m$^6$A and the amino acids residues of the aromatic cage form van der Waals (vdW) interactions and a network of hydrogen-bonds.
Furthermore, thermodynamic calculations were performed independently by \cite{li2021atomistic} and \cite{krepl2021recognition} to quantify the stabilization induced by the N6-methylation on the free energy of binding. In both cases, the calculation overestimated the stabilization of the complex compared to the experiment \cite{theler2014solution}.
To further explore the matter, Krepl \emph{et al} \cite{krepl2021recognition} also investigated the role of hydration in the binding mechanism, noticing that a water molecule was sometimes entering and leaving the binding pocket when the unmethylated adenosine was present. This water molecule occupied the location normally taken by the methyl group in presence of m$^6$A.
However, a quantiative estimate of the $\Delta \Delta G$ associate to water hydration is not trivial and was not performed in Ref.~\cite{krepl2021recognition}.
The quantitative results obtained in MD studies can also be affected by the parametrization of the employed force-field.
The N6-methyladenosine (m$^6$A) force-field parameters have been recently refined to enhance the capability of MD simulations to reproduce duplex denaturation experiments and accurately represent the populations of \emph{syn/anti} isomers of m$^6$A \cite{piomponi2022molecular}. 
Such force-field parameters are by design more reliable than those that have not been validated against experimental data
for reproducing impact of N6-methylation in the structural context of duplexes.
 Even though the destabilization induced on duplexes is generally low, it has been proven that m$^6$A can have significant impact on the structural dynamic of specific systems. For example, Jones \emph{et al} showed how the N6-methylation of adenosine favors rearrangement of nucleotides  of an RNA hairpin tetraloop \cite{jones2022structural}.
Despite this, literature mostly suggests that the primary role of m$^6$A in nature is not to alter RNA structural dynamics, but rather to facilitate RNA recognition by proteins known as m$^6$A readers.
To further improve the m$^6$A parameters, it is therefore crucial to validate them against experiments detailing the impact of N6-methylation on the free energy of binding (FEB) between RNA and m$^6$A readers.
In this work, we first investigate the influence of water displacement in and out of the aromatic cage as was observed by Krepl \emph{et al} \cite{krepl2021recognition}, and provide rigorous assessment of the significance of this process on the estimation of the binding free energies. We accomplish this by developing a protocol combining the alchemical free energy calculations (AFEC) for m$^6$A used in \cite{piomponi2022molecular} with metadynamics enhancing the displacement of water molecules both into and out of the YTH binding pocket. This new protocol leads to accurate estimations of free energy difference by sampling a variety of possible conformations of the binding pocket with respect to hydration.
Secondly, we explore the effects of the m$^6$A force-field parameters on FEB estimation.
We observe that AFEC using the parameters derived in \cite{piomponi2022molecular} does not accurately estimate the experimentally measured stabilization \cite{theler2014solution} induced by N6-methylation on the FEB between RNA and a YTH m$^6$A reader protein. We subsequently show how a better agreement with experimental results can be achieved by further refining the m$^6$A force-field using an expanded dataset compared to the one employed in \cite{piomponi2022molecular}. 
Combined with the improved precision obtained by enhancing the exploration of various hydration states within the m$^6$A binding pocket, our work comprehensively describes the stabilization induced by m$^6$A on the YTH-RNA binding free energy.

\begin{figure}
\begin{center}
\includegraphics[width=1.0\textwidth]{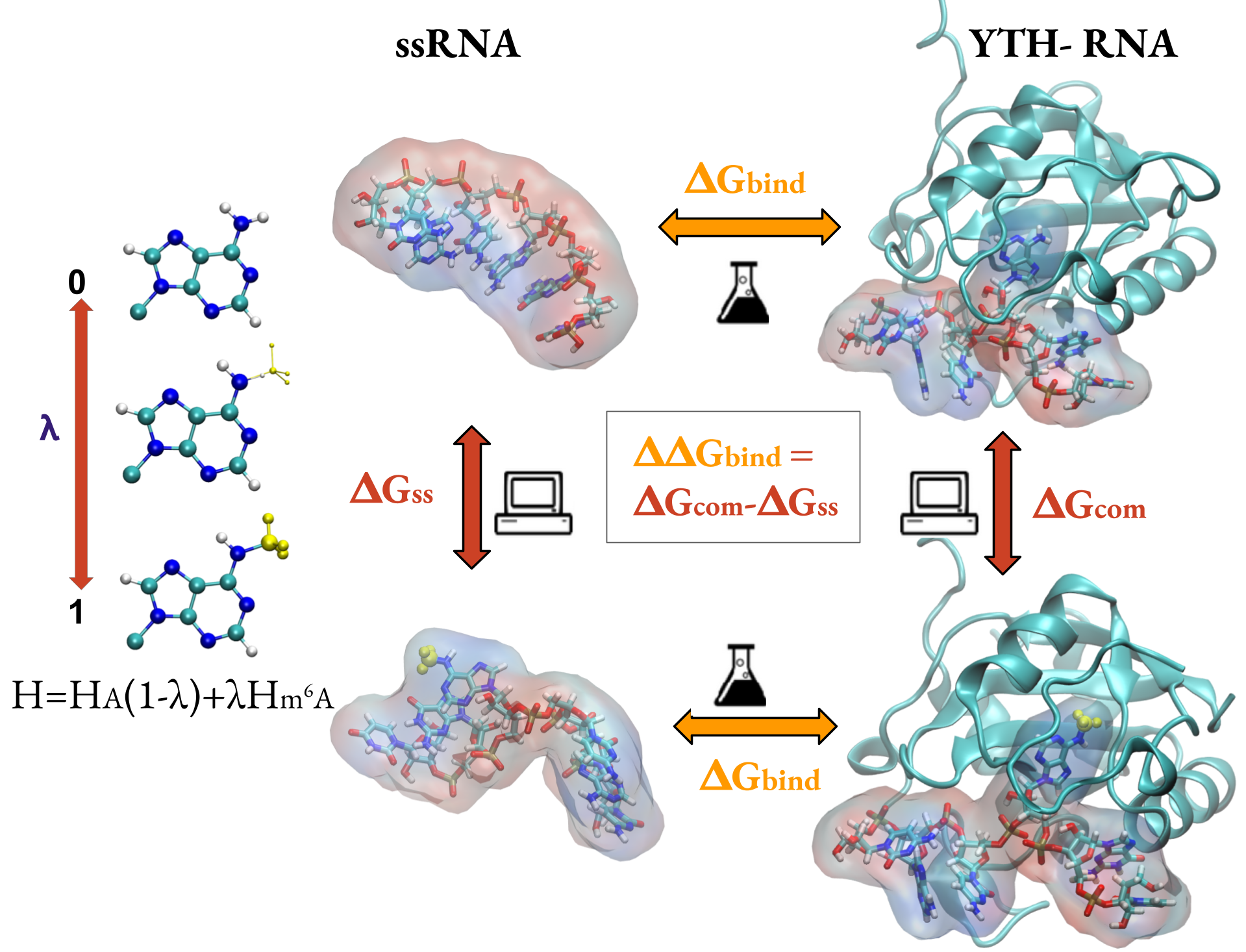}
\end{center}
\caption{Thermodynamic cycle used to compute impact of m$^6$A methylation on the FEB of the YTH-RNA complex. The relative free-energy change due to the modification can be estimated as the $\Delta \Delta G$ between AFECs performed on the complex and on the single strand RNA in solution. This quantity can be directly compared to the difference in FEB ($\Delta \Delta G_{bind}$), which was measured experimentally by Theler \emph{et al} \cite{theler2014solution}.} 
\label{yht-termo}
\end{figure}

\begin{figure}
\begin{center}
\includegraphics[width=1.0\textwidth]{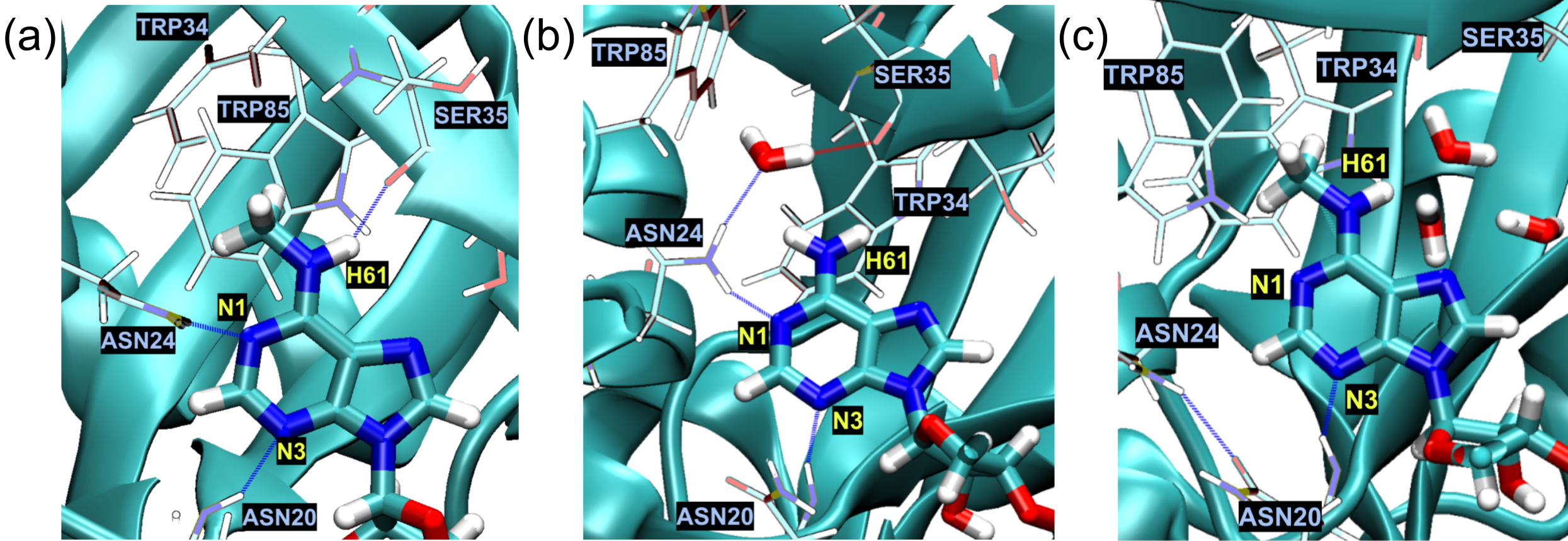}
\end{center}
\caption{Snapshots from AFEC+WT-metaD simulation representing the methylated adenosine in de-hydrated binding pocket (a), the unmethylated adenosine coordinated with a water molecule (b), or the methylated adenosine with multiple water molecules approaching the binding pocket from the H61 side (c).} 
\label{yht}
\end{figure}

\section{Material and Methods}
Starting structures for MD simulations of the YTH-RNA complex was taken from \cite{theler2014solution} (PDB ID: 2MTV) and equilibrated using the pmemd.MPI implementation of AMBER, following the same procedure used in \cite{krepl2021recognition}.
This include using the ff12SB \cite{maier2015ff14sb} 
and bsc0$\chi$OL3 (i.e., OL3)\cite{cornell1995second} \cite{perez2007refinement} \cite{zgarbova2011refinement}
force-field to describe the protein and RNA, respectively,
 the use of SPC\textbackslash E water model \cite{berendsen1987missing},
and applying HBfix potentials to increase the stability of the native A5(OPl)/LYS18(NZ) and C6(OPl)/LYS129(NZ) interactions. The only difference was in the m$^6$A parameters, where we used the fit\_A parameters derived in \cite{piomponi2022molecular}.
KCl ions \cite{joung2008determination} were added to neutralize all systems and to achieve an excess salt concentration of 0.15 M.
For the production runs, we used  a modified version of GROMACS 2020.3 \cite{abraham2015gromacs} which
also implements the stochastic cell rescaling barostat \cite{bernetti2020pressure}.
All prepared systems are listed in Section 1 of SI.

\subsection{Alchemical free energies for N6-methylation}
AFECs of the A-to-m$^6$A transformation were performed for the YTH-RNA system in order to compute $\Delta \Delta G_{bind}$. This value corresponds to the impact of N6-methylation on the YTH-RNA free energy of binding, as shown in the thermodynamic cycle depicted in Figure \ref{yht-termo}.

For the A-to-m$^6$A alchemical transformations, the protocol presented and explained in detail in \cite{piomponi2022molecular} was used. This procedure involves substituting the hydrogen H62 with a methyl group composed atoms C10, H101, H102 and H103, by gradually switching \emph{on/off} the non-bonded interactions of these atoms. We utilized the double topology scheme for the alchemical change, combined with a Hamiltonian replica exchange (HREX) with 16 replicas, in which Lennard-Jones parameters and partial charges were simultaneously interpolated.
In order to avoid singularities due to electrostatic interaction when the repulsive LJ potential is switched off \cite{mey2020best}, we used the GROMACS implemented
soft core potentials to interpolate Lennard-Jones and Coulomb potentials.

For each of the 16 replica, the systems were energy minimized and subjected to a multi-step equilibration procedure consisting of 100 ps of thermalization to 300 K in the NVT ensemble using the stochastic dynamics integrator (i.e., Langevin dynamics) \cite{goga2012efficient}, and  100 ps of pressure equilibration in the NPT ensemble using the Parrinello--Rahman barostat \cite{parrinello1981polymorphic}.
For the production runs, we used the stochastic dynamics integrator (i.e., Langevin dynamics) \cite{goga2012efficient} in combination with the stochastic cell rescaling barostat \cite{bernetti2020pressure}.
For the protein-RNA complex and ssRNA, we performed simulations of 10 and 20 ns, respectively, per replica.

\subsection{Alchemical free energies for water insertion}
We performed AFEC involving the annihilation of a water molecule in order to assess the impact of the binding pocket hydrated state found in \cite{krepl2021recognition}. This computation aims to assess:

\begin{equation}
 \Delta \Delta G^{alc-H2O}=\Delta G_{bulk}^{alc-H2O}-\Delta G_{YTH-RNA}^{alc-H2O}-\Delta G_{corr}^{alc-H2O}
\end{equation}
where $\Delta G_{bulk}^{alc-H2O}$ is associated to the annihilation of a water molecule (alc-H2O) in the bulk; $\Delta G_{YTH-RNA}^{alc-H2O}$ to the annihilation of alc-H2O in the binding pocket, restrained in the H62-coordinated position; and $\Delta G_{corr}^{alc-H2O}$ is an entropic correction accounting for the contribution of the restraint (see thermodynamic cycle in figure S2).
$ \Delta \Delta G^{alc-H2O}$ corresponds to the free energy difference between the hydrated and non-hydrated unmethylated YTH-RNA complex, and its impact on $ \Delta G_{com}$, which is the alchemical free energy difference associated to methylation in the YTH-RNA complex (see Fig.\ref{yht-termo}), can be written as follows:
\begin{equation}
\Delta  G_{com} = -k_B T \ln(e^{-\beta \Delta G_{com}^{no-H2O}} + e^{-\beta \Delta G_{com}^{H2O}}) = \Delta G_{com}^{no-H2O} -k_B T \ln(1+ e^{-\beta \Delta \Delta G^{alc-H2O}})
\end{equation}
assuming that
\begin{equation}
\Delta G_{com}^{H2O}=\Delta G_{com}^{no-H2O}+\Delta \Delta G^{alc-H2O}
\end{equation}
This assumption is based on the fact that the hydrated state is not negligible only for the $\lambda=0$ state (non-methylated A).

In these calculations we also estimated the impact of the water model in the result.
We computed $ \Delta \Delta G^{alc-H2O}$ for three different water models: SPC\textbackslash E \cite{berendsen1987missing}, TIP3P \cite{jorgensen1983comparison}, and OPC \cite{izadi2014building}.
For three different water parametrizations we performed the alchemical computation both in the YTH-RNA complex and in bulk. In the bulk simulations, all water molecules were parametrized based on the chosen model. In the YTH-RNA complex, we reparametrized only the alchemical water molecule (alc-H2O), whereas the rest of the solvent was maintained with the  SPC\textbackslash E model used in the rest of the work.
Since the alc-H2O interacts exclusively with the RNA and the YTH protein, the parametrization of the bulk of the solvent is not expected to impact these calculations.
In addition, we didn't want to include generic solvent effects on the stability of the complex, but rather to focus on the molecule directly interacting with the methyl group.
Simulations of 10 ns per replica were performed. We used 16 replica with $\lambda$ spacing: [0.00 0.01 0.03 0.05 0.10 0.20 0.35 0.45 0.55 0.65 0.80 0.90 0.95 0.97 0.99 1.00], except for $\Delta G_{com}$ with OPC and TIP3P water models where we used 8 replica,  with $\lambda$ spacing [0.00 0.03 0.06 0.13 0.30 0.50 0.75 1.00]. In $\lambda=0$, alc-H2O interactions are switched on, and \emph{vice versa}  switched off for  $\lambda=1$. The potential interpolation scheme is the same as used for the A-to-m$^6$A transformation.

During the alc-H2O AFEC in the binding pocket, a restraint was used to prevent the alc-H2O from leaving the coordination spot, in the form:
\begin{equation}
\label{restraint}
R(x)=K \mathbf{\theta(}x-c\mathbf{)}(x-c)^2
\end{equation}
where $c=0.2$ nm; $K=400$ kJ mol$^{-1}$ nm$^{-2}$ and $\theta$ is the step function. This restraint was applied on a RMSD computed on the coordinates of alc-H2O and A3 nucleobase with respect to a reference structure extracted from biased MD simulations described later.
Free energies  were computed using the binless weighted-histogram analysis method (WHAM) \cite{souaille2001extension,shirts2008statistically,tan2012theory}. In the YTH-RNA case, the $\Delta G_{YTH-RNA}^{alc-H2O}$ accounts for switching from the coupled and unrestrained alc-H2O to the decoupled and restrained alc-H2O (see Fig. S2).
In the bulk simulations, the alchemical water is always unrestrained. As a consequence, an entropic correction is necessary to keep the relationship between the standard state volume and the accessible space in the binding pocket \cite{tanida2020alchemical}. 
This correction $\Delta G_{corr}^{alc-H2O}$ was evaluated assuming that the bias potential (\ref{restraint}) restrains the alc-H2O in a volume ($V_{acc}$) that can be derived through a numerical integration, following the expression:

\begin{equation}
V_{acc} = \int e^{-\beta R(x)} dx = \frac{4 \pi c^3}{3} + \int_c^{\infty} 4\pi r^2 e^{- \beta K (r-c)^2} dr 
\end{equation} 

The entropic correction was finally evaluated as follows:

\begin{equation}
\Delta G_{corr}^{alc-H2O} = - K_b T \ln(\frac{V_{acc} N}{V_{box}})
\end{equation} 

where $N/V_{box}$ is the density of water molecules in the bulk.

\subsection{Metadynamics}

In order  to accelerate the process of water exchange in and out the YTH binding pocket, we combined the A-to-m$^6$A AFEC with a WT-MetaD \cite{laio2002escaping,barducci2008welltempered,bussi2020using} acting on a collective variable (CV) which is able to quantify the number of water molecules approaching the binding pocket. 
We thus chose as biased CV the coordination number between two groups of atoms as implemented in PLUMED \cite{tribello2014plumed}
using the standard switching function:
\begin{equation}
\label{cn}
CN = \sum _{i \in A} \sum _{j \in B} \frac{1}{1 + (\frac{r_{ij} }{r_0})^{6}}
\end{equation} 
In our implementation, we define group $A$ as a single point at the center of the atoms N6; H61; and C10, whereas group $B$ includes all water oxygens in the system. $r_0$ was set to 0.45 nm. 
A WT-MetaD on $CN$ without any restraints could cause multiple water molecules entering the binding pocket at the same time, likely causing the RNA to unbind. To prevent this, we
included an upper harmonic wall potential, defined as follows:
\begin{equation}
\begin{cases}
	V_{walls}(x_i)= K (CN(x_i) - UW)^2  \; \;  ;  \; \; \; CN(x_i) > UW \\
	V_{walls}(x_i)= 0  \; \;  ;  \; \; \; CN(x_i) \leq UW 
\end{cases}
\end{equation}
where we set $K=200$ kJ/mol and $UW=2.5$.
The metadynamics was performed using the PLUMED package \cite{tribello2014plumed}, depositing a Gaussian every 500 time steps, with initial height equal to 5 kJ/mol and width $\sigma = 0.05$. The bias factor was set to 3.
The calculation of $CN$ was accelerated making use of a neighbor list, which makes it that only a relevant subset of the pairwise distance are calculated at every step. We used a neighbor list cut-off of 0.8 nm, updating the lists every 10 steps.
We first performed the AFEC computation with WT-MetaD on $CN$ running for 20 ns per replica.
We then performed another AFEC with 100 ns per replica with a static bias, by restarting the previous AFEC with WT-Metad without further updating the bias,
as it is done in metadynamics with umbrella-sampling corrections \cite{babin2006free}. In the following, we only analyze the results obtained from the static bias simulations.

\subsection{Hamiltonian Replica Exchange on m$^6$A charges}
\label{hrex}

To compute $\Delta \Delta G_{bind}$ for alternative m$^6$A charges parametrization with respect to the one used in AFEC simulations (fit\_A), we implemented new sets of simulations using an Hamiltonian Replica Exchange (HREX) scheme similar to the one used in AFEC, but where the initial and final states in the integration correspond to different parametrization of the m$^6$A charges.
$\lambda =0$ would correspond to methylated state with fit\_A charges, whereas $\lambda =1$ would correspond to the methylated state with alternative charges parametrization. In particular, we considered in this work charges parametrization for m$^6$A published by Aduri \emph{et al} \cite{aduri2007amber} and by Krepl \emph{et al} \cite{krepl2021recognition}.
By performing this transformation on the YTH-RNA complex and on the corresponding ssRNA, and computing respectively $\Delta G_{com}^{\Delta Q}$ and $\Delta G_{ss}^{\Delta Q}$,  we can compute the  $\Delta \Delta G_{bind}^{ff}$ for the two different force fields as:
\begin{equation}
\Delta \Delta G_{bind}^{ff} = \Delta \Delta G_{bind}^{fit\_A} + \Delta \Delta G_{bind}^{\Delta Q}
\end{equation}
where
\begin{equation}
\Delta \Delta G_{bind}^{\Delta Q} = \Delta G_{ss}^{\Delta Q} -\Delta G_{com}^{\Delta Q}
\end{equation}

We performed simulations starting from the YTH-RNA complex and the ssRNA  (5$^{\prime}$- CGm$^6$ACAC-3$^{\prime}$), using the HREX scheme with only 2 replicas for the fit\_A-to-Aduri integration, and 4 replicas for the fit\_A-to-Krepl integration. This choice of number of replicas allows ensuring averaged transition probabilities over 20\%. Simulations were 10 ns per replica long. Free energies difference were computed with BAR method implemented in GROMACS \cite{bennett1976, wu2005bar}.

\subsection{Force-field fitting}
We refined the m$^6$A force-field by re-applying the fitting procedure described in our previous work \cite{piomponi2022molecular}, which involves adjusting a subset of the partial charges and a dihedral potential acting on the torsional angle $\eta _6$ identified by atoms N1--C6--N6--C10. This procedure allows to use the A-to-m$^6$A AFECs as a reference to match experimental data. %
Here, the protocol was readapted in order to allow the fitting of different sets of charges, involving the minimization of a cost function defined as:
\begin{equation}
\label{costfunc}
C = \chi ^2 + \alpha \sum_{i=0}^N \Delta Q_i^2 + \beta V_{\eta}^2 = \chi ^2 + \alpha [\sum_{i=1}^N \Delta Q_i^2 + (\sum_{i=1}^N \Delta Q_i)^2] + \beta V_{\eta}^2
\end{equation} 
where $N$ is the number of fitted charges minus one and $\chi ^2$ measures the discrepancy between computations and experiments as follows:
\begin{equation}
\chi ^2 = \frac{1}{N_{exp}} \sum_{j=1}^{N_{exp}} \frac{(\Delta \Delta G_j^{AFEC} - \Delta \Delta G_j^{exp})^2} {\sigma _j^2}
\label{eq_chi}
\end{equation}
Here $\sigma _j$ correspond to the experimental errors. The regularization terms on the charges and the torsional $\eta _6$ are governed by the hyperparameters $\alpha$ and $\beta$ and are needed to avoid overfitting on the training set.
$\Delta \Delta G$s for perturbed parameters were computed through reweighting, by considering that the potential energy change associated with charges and torsion perturbation is:
\begin{equation}
\label{DeU}
\Delta U(x)=\sum_{i=N}^5 K_i(x) \Delta Q_i + \\ \sum_{i=1}^N\sum_{j=i}^N K_{ij}(x) \Delta Q_i \Delta Q_j + V_{\eta} [1 + cos(\eta_6 (x_i) - \pi)]
\end{equation}
In total, for every analyzed snapshot ($x$), $N(N-1)/2 + 2N$ coefficients ($K_i$ and $K_{ij}$) can be precomputed that allow obtaining the energy change
for arbitrary choices of $\Delta Q$ with simple linear algebra operations, without the need to recompute electrostatic interactions explicitly.
These coefficients were obtained by using GROMACS in rerun mode for $N(N-1)/2 + 2N$ sets of test charge perturbation,
which were extracted from a Gaussian with zero average and standard deviation set to 1 \emph{e}.

In order to asses the reliability of the $\Delta \Delta G$ estimations obtained through reweighting, we quantify the statistical significance of our estimation by looking a the Kish Size Ratio (KSR), defined as:

\begin{equation}
KSR=\frac{KS_{\lambda=1}}{KS^0_{\lambda=1}}
\end{equation}
where
\begin{equation}
KS_{\lambda=1} = \frac{[\sum_x w(x) e^{-\beta(\Delta E(x)+\Delta  U(x))}]^2}{\sum_x [w(x) e^{-\beta (\Delta E(x)+\Delta  U(x))}]^2}
\end{equation}
is the Kish effective sample size \cite{GrayKish1969,rangan2018determination} of the ensemble obtained with perturbed parameters, whereas:
\begin{equation}
KS^0_{\lambda=1} = \frac{[\sum_x w(x) e^{-\beta \Delta E(x)}]^2}{\sum_x [w(x) e^{-\beta \Delta E(x)}]^2}
\end{equation}
is the Kish effective sample size of the unperturbed ensemble.

\section{Results}

We here report the results of a computational study aimed at investigating the role of m$^6$A modifications in RNA-protein interactions,
with honest assessment of the impact of limited sampling and details of force-field parametrizations.
The $\Delta G$s associated with methylating the adenosine in the YTH-RNA complex and in the corresponding ssRNA are shown in Table \ref{pinolo}.
The corresponding $\Delta \Delta G_{bind}$, which reports on the stabilization effect of the methylation on the complex, is predicted to be 22.1 $\pm$ 0.8 kJ/mol.
This result is a signficiant overestimation of the experimental value (9.9 $\pm$ 0.1 kJ/mol) \cite{theler2014solution},
and is even larger than the estimation reported in \cite{krepl2021recognition} (18.0 kJ/mol), where a different parametrization for m$^6$A was used.
One can expect the computational overestimation of $\Delta \Delta G_{bind}$ to be mainly caused by inaccuracy of the force-field and
limited sampling.
In Krepl parametrization, the negative partial charges of nitrogens N1 and N3 have lower absolute values compared to the fit\_A counterparts (see Fig. \ref{charges_scheme} and Table S2). Both N1 and N3 form hydrogen bonds in the binding pocket (see Figure \ref{yht}), which is stronger in the fit\_A case, possibly explaining why we observe a larger stabilization of the complex induced by the N6-methylation.
As for the limited sampling issue, as suggested previously by \cite{krepl2021recognition}, a factor which could impact the precision of $\Delta \Delta G_{bind}$ calculation is the role of hydration in the binding pocket. In our unbiased AFEC computation we never observe water molecules entering the binding pocket, but a variety of different hydrated states could be ideally explored in fully converged simulations. 

We remark the fact that we simulate the ssRNA with SPC\textbackslash E water model to be consistent with \cite{krepl2021recognition}. However, in \cite{piomponi2022molecular}, we simulated the ssRNA in solutions using TIP3P water model. To check the consistency of the results with respect to different water models, we additionaly performed a simulation of the ssRNA using TIP3P. We obtained a $\Delta G$ of 206.1 $\pm$ 0.4 kJ/mol, which is consistent, within the statistical error, with the one obtained using SPC\textbackslash E (205.2 $\pm$ 0.7 kJ/mol).

\begin{table}

\begin{center}
\resizebox{\columnwidth}{!}{%
\begin{tabular}{|l|c|c|c|c|}
\hline
\multicolumn{1}{|c|}{\textbf{}}            &            \multicolumn{2}{ |c| }{\textbf{fit\_A}}            &\multicolumn{2}{c|}{\textbf{fit5\_AC}}  \\
\hline
   &           $\Delta G$         & $\Delta \Delta G$  &     $\Delta G$            & $\Delta \Delta G$ \\
\hline
ssRNA & 205.2 $\pm$ 0.7 & 0 & 237.0 $\pm$ 0.6 & 0 \\
\hline
YTH-RNA  &  183.1 $\pm$ 0.4   & 22.1 $\pm$ 0.8 & - & - \\ 
\hline
YTH-RNA  + MetaD & 185.4 $\pm$ 1.3   & 19.7  $\pm$  1.5 &  224.6 $\pm$ 1.0 &  12.4  $\pm$ 1.2 \\
\hline

\end{tabular}%
}
\caption{Free energies differences computed through the A-to-m$^6$A AFEC with different parametrizations, reported in kJ/mol.
For the YTH-RNA complex, $\Delta G$s were also estimated by implementing the AFEC with a WT-metaD accelerating the water displacement in the binding pocket. Results are given for the unbiased system (YTH-RNA),
 and the system affected by the static bias produced by a previous WT-metaD (YTH-RNA  + MetaD).
 The impact of N6-methylation on the FEB is computed as $\Delta \Delta G_i  = \Delta G_{ss} - \Delta G_i$.}
\label{pinolo}
\end{center}

\end{table}

\subsection{Alchemical free energies for water insertion}
As suggested by Krepl \emph{et al} al \cite{krepl2021recognition}, one of the factors contributing to the overestimation of $\Delta \Delta G_{bind}$ could be the omission of scenarios where a water molecule is situated inside the binding pocket and coordinates with atom H62. Such a configuration is plausible at $\lambda=0$ but becomes improbable at $\lambda=1$ due to steric hindrance from the methyl group. In plain MD simulations it was observed that a water molecule stays inside the binding pocket 10\% of the time when standard adenosine is bound \cite{krepl2021recognition}.
This suggests that the hydration plays a minor role in $\Delta \Delta G_{bind}$. To investigate this further, we conducted AFEC simulations involving the annihilation of a water molecule within the binding pocket.
All computed free energies for the alchemical transformation of water are detailed in Table \ref{tableDeGwater}. Notably, all $\Delta G^{alc-H2O}$s values are positive, indicating a disfavoring of the hydrated state, which aligns with our expectations. Intriguingly, when using the TIP3P and OPC water models, the hydrated state becomes even more disfavored, resulting in a further marginal impact on $\Delta \Delta G^{bind}$. 
We can quantify the correction to $\Delta \Delta G^{bind}$, in relation to the estimates obtained in the previous section that did not account for hydration effects, as $\Delta \Delta \Delta G_{bind} = - k_B T \ln(1+ e^{-\beta \Delta \Delta G^{alc-H2O}})$. These corrections are presented in the fourth column of Table \ref{tableDeGwater}, and they are found to be very small in comparison to the differences between the experimental  values and those estimated in computational studies.
Therefore, we conclude that this hydrated configuration has only a very minor impact on the FEB and cannot account for the mismatch between experimental and computational $\Delta \Delta G^{bind}$ data.

\begin{table}

\begin{center}
\resizebox{0.8\columnwidth}{!}{%
\begin{tabular}{|l|c|c|c|c|}

\hline

\hline
   &           $\Delta G_{bulk}^{alc-H2O}$         &  $\Delta G_{com}^{alc-H2O}$            & $\Delta \Delta G^{alc-H2O}$ & $\Delta \Delta \Delta G_{bind}$ \\
\hline
SPC\textbackslash E & 29.51 $\pm$ 0.08 &  22.1 $\pm$ 0.6 & 4.8 $\pm$ 0.6  & -0.33 \\
\hline
TIP3P  &  25.48 $\pm$ 0.07   & 15.4 $\pm$ 1.0 &  7.5 $\pm$ 1.0 & -0.12\\ 
\hline
OPC & 33.7 $\pm$ 0.5 & 23.8 $\pm$ 1.2  & 7.3 $\pm$ 1.2 & -0.13    \\
\hline

\end{tabular}%
}
\caption{Free energy differences computed through AFEC involving the annihilation of a water molecules in bulk ($\Delta G_{bulk}^{alc-H2O}$) or coordinated to the adenosine amino group in the in the YTH-RNA complex ($\Delta G_{com}^{alc-H2O}$). The results are given for different water models, and are reported in kJ/mol. $\Delta \Delta G^{alc-H2O}$ measures the preference for the water molecules to stay in the bulk rather than inside the binding pocket. $\Delta \Delta \Delta G_{bind}$ estimates the correction induced to the $\Delta \Delta G_{bind}$ by taking into account the specific hydrated state considered here.}
\label{tableDeGwater}
\end{center}
\end{table}

\subsection{Enhancing binding pocket water exchange in alchemical simulation}
The hydrated state considered in previous section is only one of the possible metastable states, individuated from plain MD simulations \cite{krepl2021recognition}, but in principle different hydrated states of the binding pocket might occur.
Therefore, we aimed to improve the precision of our AFEC, by allowing more exhaustive sampling with respect to water movement in and out of the binding pocket.
To that end, we performed new AFEC of the YTH-RNA complex coupled with a WT-metaD acting
on a coordination switching function $CN$ (see Eq. \ref{cn}).
Fig. S4 shows the values of $CN$ and a control variable $d$ along the demuxed continuous trajectories. $d$ is defined as a distance between the center of mass of m$^6$A nucleobase and the center of the residues forming the binding pocket.  This variable can be monitored to check that the RNA does not exit from the binding pocket.
During the static bias simulations, the hydrated state described in previous section (a water molecule coordinated with atom H62 of m$^6$A) appears only in two replicas:
In one of them it is always present and as a consequence this trajectory is unable to fully explore the $\lambda$ ladder (see top left corner of Fig. S1), due to the steric clashes between the water molecule and the appearing methyl group.
In the second  trajectory, there is initially no water present inside the binding pocket, but one enters after more than 80 ns and does not leave the pocket again. This once again prevents exploration of high $\lambda$ values. In all other trajectories, the cases of $CN$ going to high values correspond to multiple water molecules approaching the binding pocket, but remaining stuck on the other side of the amino group, coordinating with atom H61 and residue SER35, as shown in the snapshot in Fig. \ref{yht}c.
Although this enhanced sampling simulation shows many limitations, such as the inability to transition in both directions, the obtained sampling is more exhaustive than the one obtained previously without biasing the $CN$.
Consequently, the $\Delta G$ computed with WHAM from the biased simulation is 185.4 $\pm$ 1.3 kJ/mol (see Table \ref{pinolo}), resulting in a  $\Delta \Delta G$ of 19.7 $\pm$ 1.5 kJ/mol
, which is  slightly reduced compared to the estimation done without enhancing the water displacement ($22.1 \pm 0.8$ kJ/mol), but still  overestimated compared to the experimental reference (9.9 kJ/mol).

\subsection{Exploring m$^6$A force-field perturbation effects on FEB}
\begin{figure}
\begin{center}
\includegraphics[width=1.0\textwidth]{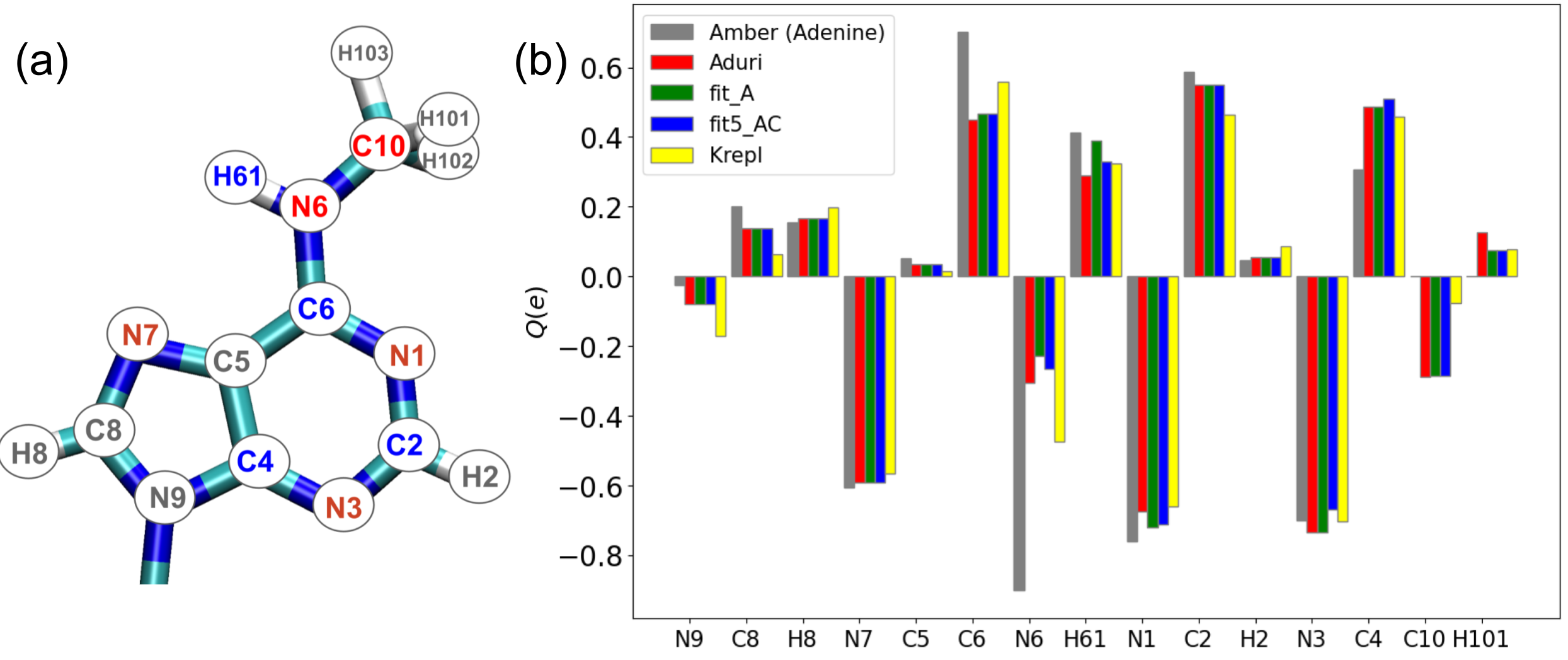}
\end{center}
\caption{(a) m$^6$A nucleobase scheme. Atoms are colored based on partial charges tendency. (b) Partial charges for different parametrizations of adenosine and m$^6$A.}
\label{charges_scheme}
\end{figure}

Since water hydration appears to have a limited impact on the accuracy and precision of free energy estimations, we tested the hypothesis that the primary reasons for the discrepancies between experimental and computational results stem from the inaccuracies in the force-field parameters. We have used here the fit\_A force-field for m$^6$A, that we derived in our previous work \cite{piomponi2022molecular} by refining the Aduri force-field \cite{aduri2007amber}  to better match \emph{syn/anti} populations and duplex denaturation experiments. This refinement involved adjusting a subset of partial charges that play a significant role in the stability of duplexes, particularly with respect to hydrogen bond strength involving  WC edge atoms of the nucleobase. As far as the m$^6$A recognition by the YTH protein is concerned, there are other atom parameters which may play significant role in the stabilization. For instance, in the YTH complex m$^6$A performs hydrogen bonding with the protein residues also on its sugar edge, so it could be useful to refine the partial charge of nitrogen atom N3, which was not considered in the fit\_A fitting. 
In \cite{piomponi2022molecular} we have demonstrated how small variations in the partial charges can have significant impact on free energy differences induced by small chemical modifications such as the methylation. 
To compare the performance of different parameters in the context of the YTH-RNA, we first compute $\Delta \Delta G_{bind}$ for m$^6$A force-field alternative to fit\_A,
Aduri \cite{aduri2007amber} and Krepl \cite{krepl2021recognition}, using an HREX scheme described in Section \ref{hrex}. The results of these computations are listed in Table 1 of SI, whereas the relative $\Delta \Delta G_{bind}$ are depicted in Fig. \ref{fit5_recap}.
Not surprisingly, Krepl and Aduri force-field destabilize the YTH-RNA complex with respect to Fit\_A. Indeed, the latter parametrization is characterized by atom H61, N3 and N1 being more polar than in the other cases, as shown in Fig. \ref{charges_scheme}. All these atoms form hydrogen bonds in the aromatic cage, respectively with SER35, ASN20 and ASN24 residues, as shown in Fig.\ref{yht}. 
Based on our estimations of $\Delta \Delta G_{bind}$s, the Aduri force-field seems to be the most compatible with the experimental values, as it can be seen in the plot in Fig. \ref{fit5_recap}a.
However, the Aduri force-field is not able to reproduce denaturation experiments and the \emph{syn/anti} populations, as shown in \cite{piomponi2022molecular}.
This leads to the conclusion that
none of the so far explored m$^6$A force-field is able to simultaneously reproduce isomer populations and denaturation experiments of duplexes as well as calorimetry experiments on the YTH-RNA complex. 
Before resorting to fitting charges, we also
investigated if perturbations in the LJ parameters of the methyl group hydrogens could have impact on the $\Delta \Delta G_{bind}$ (see Fig. S3). We observed that reasonable perturbations on these LJ parameters do not lead to sufficient improvements.

\subsection{Force-field refinement}
As demonstrated above, the m$^6$A fit\_A force-field fails to accurately reproduce the results of ITC experiments on the YTH-RNA complex. 
Consequently, we decided to refine the m$^6$A parametrization further by extending the fitting procedure outlined in \cite{piomponi2022molecular}. This extension involves incorporating an expanded experimental dataset, which includes the YTH-RNA $\Delta \Delta G_{bind}$. The list of experiments considered for this fitting is provided in Table \ref{table-yht} and is divided into a training dataset and a validation dataset.
The fitting procedure presented in \cite{piomponi2022molecular} was re-adapted to work over the simulations performed with fit\_A parametrization on systems A1-A2-A3-A4-A5, along with the YTH-RNA $\Delta \Delta G_{bind}$, which we will refer to as the C1 system.
Systems B1-B2-B3-B4-B5, in addition to the $\Delta G_{syn/anti}$ for system A2 (A2$_{syn/anti}$), were instead used to validate the parametrization derived from the fitting process.
We have also refined the torsional parameter $V_{\eta}$ and, in conjunction with it, we aimed to optimize two distinct subsets of partial charges independently, resulting in the creation of two separate parametrizations, namely
fit6\_AC (fit on atoms C6-N6-H61-N1-C10-H101 partial charges); and
fit5\_AC (fit on atoms N6-H61-N1-N3-C4 partial charges).
While fit6\_AC aims to explore the same  charges space explored by the fittings illustrated in \cite{piomponi2022molecular}, fit5\_AC is designed to investigate a smaller multidimensional space that includes atoms N3 and C4. In particular, the polarity of N3 may play a significant role in stabilizing the binding pocket in C1, as this atom forms hydrogen bonds with the ASN20 residue within the aromatic cage. Additionally, we have included in the set the charges of N1 and H61 atoms, which are involved in hydrogen bonding both in the dsRNA (A2--A4 and B1--B5) and in the YTH binding pocket (C1). Therefore, we expect that the fitting process would be highly sensitive to these charges. Atoms N6 and C4 are primarily intended to absorb the perturbations introduced by the fitting process of the other three charges.

\begin{table}

\begin{center}
\resizebox{\columnwidth}{!}{%
\begin{tabular}{|l|c|c|l|c|c|}
\hline
\multicolumn{3}{|c|}{\textbf{Training Set}}            &            \multicolumn{3}{ |c| }{\textbf{Validation Set}}     \\
\hline
   System      & $\Delta \Delta G$ (kJ/mol)  &     Exp            & System      & $\Delta \Delta G$ (kJ/mol)  &     Exp  \\
\hline
A1$_{syn/anti}$ & 6.3 $\pm$ 0.5 & NMR \cite{roost2015structure} & A2$_{syn/anti}$ & -11 $\pm$ 2 & NMR \cite{liu2020quantitative} \\
\hline
A2 & 1.7 $\pm$ 0.9 & DE \cite{roost2015structure} & B1 & 2.5 $\pm$ 2.1 & DE \cite{kierzek2022secondary} \\
\hline
A3 & 7.1 $\pm$ 0.9 & DE \cite{roost2015structure} & B2 & 2.1 $\pm$ 1.3 & DE \cite{kierzek2022secondary} \\
\hline
A4 & -2.5 $\pm$ 1.2 & DE \cite{roost2015structure} & B3 & 5.4 $\pm$ 1.3 & DE \cite{kierzek2022secondary} \\
\hline
A5 & -1.7 $\pm$ 0.9 & DE \cite{roost2015structure} & B4 & 8.6 $\pm$ 0.8 & DE \cite{kierzek2022secondary} \\
\hline
C1 & 9.9 $\pm$ 0.5 & ITC \cite{theler2014solution} & B5 & 1.7 $\pm$ 1.0 & DE \cite{kierzek2022secondary} \\
\hline
\end{tabular}%
}
\caption{List of systems and relative experimental $\Delta \Delta G$ considered in the fitting. These values represent m$^6$A rotamer preference (A1/A2$_{syn/anti}$); destabilization induced by the N6-methylation on dsRNAs (A2-A5, B1-B5); and impact of the N6-methylation on the FEB of the YTH-RNA system (C1). $\Delta \Delta G$s and relative error are derived from nuclear magnetic resonance (NMR) experiments, optical melting denaturation experiments (DE) and  isothermal Titration calorimetry (ITC) measurements, as indicated.}
\label{table-yht}
\end{center}

\end{table}

\begin{figure}
\begin{center}
\includegraphics[width=1.0\textwidth]{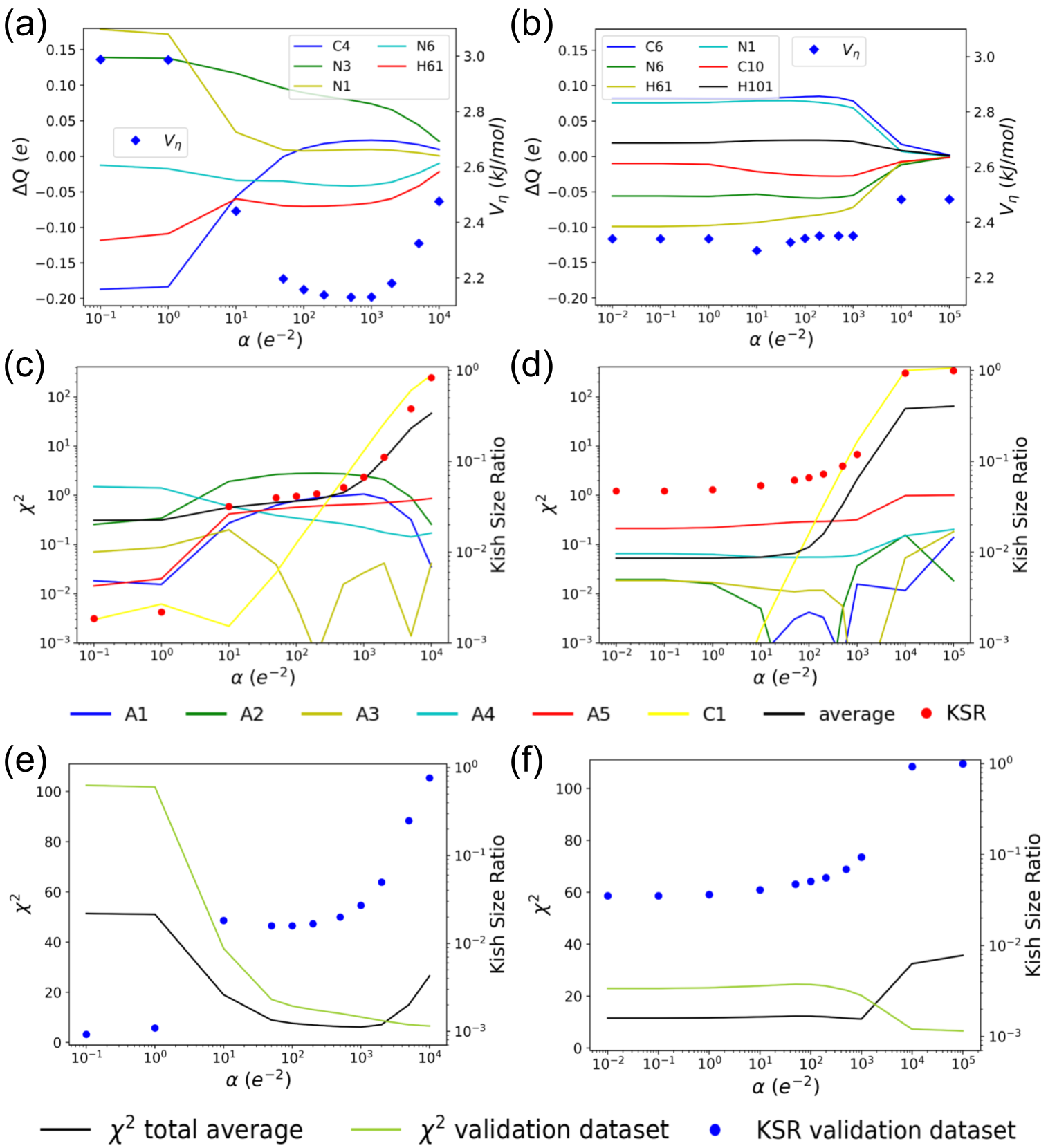}
\end{center}
\caption{Results of the charge fitting procedure. Parameters ($\Delta Q$ and $V_{\eta}$) obtained fitting on the traning dataset
as a function of $\alpha$, with $\beta=0$, for fit5\_AC (panel a) and fit6\_AC (panel b).
$\chi^2$ errors for individual experiments of the training dataset
and Kish size ratio (KSR)
as a function of $\alpha$, with $\beta=0$,  for fit5\_AC (panel c) and fit6\_AC (panel d).
Averaged $\chi^2$ obtained for the total dataset (black line) and on the validation dataset (yellow line),  for fit5\_AC (panel e) and fit6\_AC (panel f). The KSR computed on the validation dataset is also shown (blue dots). } 
\label{fits_AC}
\end{figure}

The results of the two fittings are shown in Fig. \ref{fits_AC}. 
Based on the insights learned from the cross validations performed in our previous fitting \cite{piomponi2022molecular}, we know that regularization on the charges is necessary to avoid overfitting, whereas regularization on $V_{\eta}$ can be discarded by setting $\beta=0$ in Eq. \ref{costfunc}.
Panels \ref{fits_AC}a and \ref{fits_AC}b display the optimized parameters  at different $\alpha$ values, while panels \ref{fits_AC}c and \ref{fits_AC}d depict the corresponding $\chi ^2$ values and KSR values for each parameter set obtained at different $\alpha$ values. In both cases, at lower $\alpha$ values, the fitting effectively enforces experiment C1, at the expense of yielding very low KSR values, hence making the free energy estimation not statistically significant. As $\alpha$ values increase, the $\chi ^2$ values for C1 rise significantly, while the $\chi ^2$ values for other experiments remain relatively stable and sometimes even decrease. This outcome is not unexpected, as higher values of $\alpha$ constrain the parametrization to the fit\_A force-field, which was designed to match the A1--A5 experimental data and is intended to perform well for them.
The minimum $\alpha$ values that ensure a KSR above 0.1 are respectively $\alpha=1000$ e$^{-2}$ and $\alpha=2000$ e$^{-2}$ for fit6\_AC and fit5\_AC. The charge values obtained by minimizing the cost function for these $\alpha$ values were selected as the results of the two fittings. This choice is further validated by the estimation of the $\chi ^2$ on the validation dataset, as shown in panels  \ref{fits_AC}e and  \ref{fits_AC}f.
$\Delta Q$s values with respect to fit\_A parametrizations are shown in Table \ref{tablePar_AC}, along with the $V_{\eta}$ values.
It's worth noting that both fittings result in a decrease in the polarity of atoms N1, H61, and also N3 in the fit5\_AC case, as expected.

Table \ref{table_fitAC_chis} provides the averaged $\chi ^2$ values computed separately for the training and validation datasets, as well as the overall average. The columns labeled fit6\_AC (rew) and fit5\_AC (rew) represent the results obtained through reweighting. While fit6\_AC demonstrates better performance in the training dataset compared to fit5\_AC, it exhibits poor performance in the validation dataset, resulting in a total $\chi ^2$ score that is even worse than the initial state of the fitting (fit\_A). On the other hand, Fit5\_AC performs quite well on the validation dataset, making it the better candidate to serve as the best parametrization to align with the entire dataset.
Based on this observation, we conducted new simulations of the complete dataset shown in Table \ref{table-yht} using the fit5\_AC m$^6$A force-field. This also included a new simulation of the YTH-RNA complex employing the same AFEC+WT-MetaD procedure as previously utilized (results reported in Table \ref{pinolo}).
All the computed alchemical $\Delta G$s for systems A and B related to this work are listed in table S3. 

\begin{figure}
\begin{center}
\includegraphics[width=1.0\textwidth]{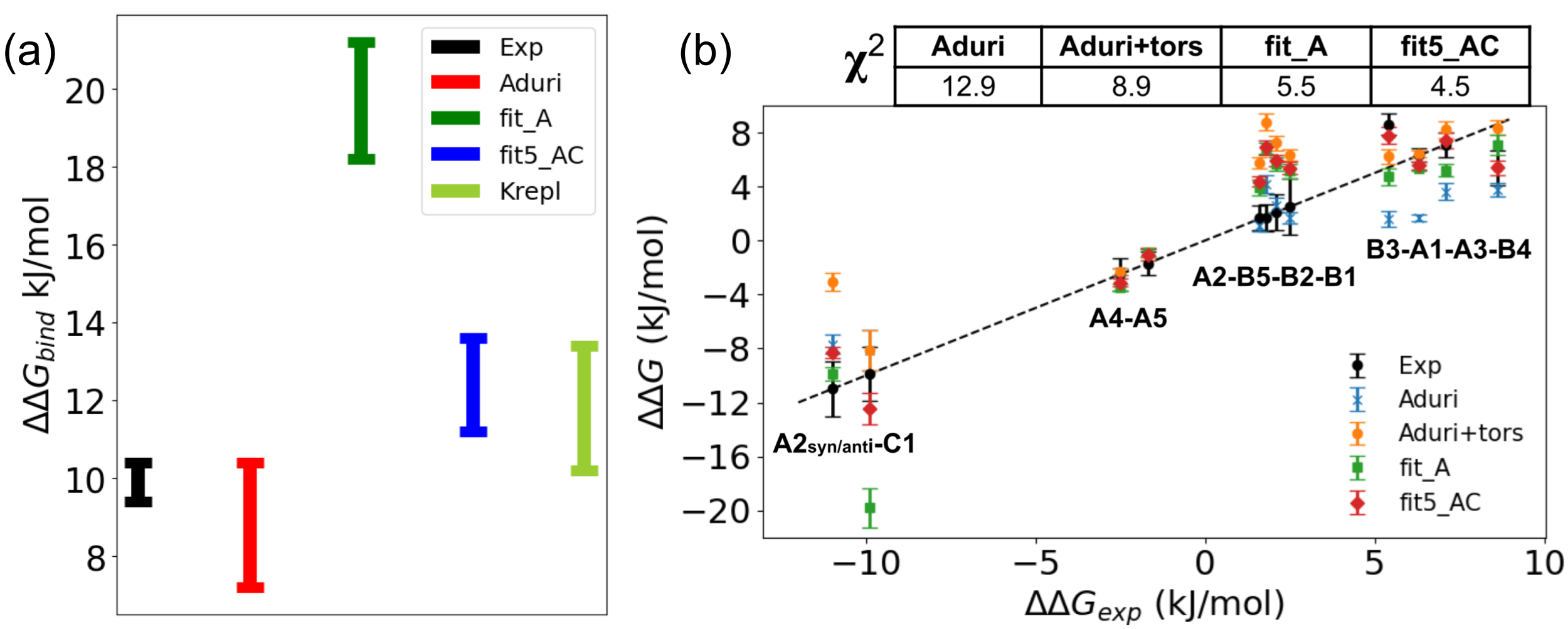}
\end{center}
\caption{(a) $\Delta \Delta G_{bind}$ values and relative experimental or statistical error  (b) $\Delta \Delta G$s computed for each of the 12 analyzed systems with 4 different sets of parameters. $\Delta \Delta G$ for system C1 is shown as the inverse of $\Delta \Delta G_{bind}$.
$\chi ^2$ obtained for each force-field set of parameters are shown in the table.}
\label{fit5_recap}
\end{figure}

Panel \ref{fit5_recap}a presents a summary of all the $\Delta \Delta G_{bind}$ values computed in this study for different parametrizations, alongside the experimental value. 
The newly fitted parameters (fit5\_AC) do not accurately
replicate the experimental $\Delta \Delta G_{bind}$ as effectively as the other alternative parameters settings. However, they strike a balanced compromise between
matching $\Delta \Delta G_{bind}$ and the other experimental $\Delta \Delta G$ values within our dataset, as demonstrated in panel \ref{fit5_recap}b.
Specifically, while A2-A3 and B1--B5 demand an enhancement in the polarity of N1 and H61 atoms to strengthen hydrogen bonds and stabilize the duplexes, the C1 experiment necessitates the opposite effect to reduce the overestimation of $\Delta \Delta G_{bind}$.
In comparison to fit6\_AC, fit5\_AC exhibits greater flexibility by allowing adjustments to the partial charge of atom N3, which is believed to be more sensitive to experiment C1 than in the duplex systems, where N3 does not form hydrogen bonds.

\begin{table}
\begin{center}
\resizebox{\columnwidth}{!}{%
\begin{tabular}{|l|c|c|c|c|c|c|c|c|r|}
\hline
      & \textbf{C6} (e) & \textbf{N6} (e) & \textbf{H61} (e) & \textbf{N1} (e) &\textbf{C10} (e) & \textbf{H100} (e) & \textbf{N3} (e) & \textbf{C4} (e) & \textbf{$V_{\eta}$} (kJ/mol)\\
\hline
\textbf{fit5\_AC} & 0 & -0.0363 & -0.0595 & 0.0086 & 0 & 0 & 0.0657 & 0.0215 & 2.18 \\
\hline
\textbf{fit6\_AC} & 0.0644 & -0.0550 & -0.0720 & 0.0687 &  -0.0272 &  0.0211  & 0  & 0 & 2.35\\
\hline
\end{tabular}%
}
\caption{Charge modifications ($\Delta Q$s) and torsional potential ($V_{\eta}$) for the fitting performed on the training data set AC.} 
\label{tablePar_AC}
\end{center}
\end{table}

\begin{table}

\begin{center}
\resizebox{\columnwidth}{!}{%
\begin{tabular}{|c|c|c|c|c|c|c|}
\hline
$\chi ^2$ & Aduri & Aduri$+$tors & fit\_A & fit6\_AC (rew) & fit5\_AC (rew) & fit5\_AC \\
\hline
Training Set & 16 & 3.8 & 4.5 & 0.33 & 0.9 & 2.2 \\
\hline
Validation Set & 9.7 & 14 & 6.5 & 18 & 7.5 & 6.7 \\
\hline
Total & 12.9 & 8.9 & 5.5 & 9 & 4.2 & 4.5 \\
\hline
\end{tabular}%
}
\caption{Results of the fitting. $\chi ^2$ computed for the training data set AC (second row), the validation data set B$+$A2$_{syn/anti}$ (third row), and the total average (fourth row) for different m$^6$A force-field.  fit6\_AC (rew) and fit5\_AC (rew) are $\chi ^2$ values obtained through the fitting and computing free energies by reweighting.}
\label{table_fitAC_chis}
\end{center}

\end{table}

\section{Conclusions}

In this work we explored the role of m$^6$A in RNA recognition, in the context of the YTH domain of the YTHDC1 protein. 
MD simulations have already been used to investigate the binding mechanism in this system \cite{li2019flexible, li2021atomistic, krepl2021recognition}, but they failed to accurately reproduce the m$^6$A-induced stabilization ($\Delta \Delta G_{bind}$) of the YTH-RNA binding as expected from experiments \cite{theler2014solution}.
We here investigated the possible reasons for these discrepancies, namely insufficient sampling and incorrect force-field parametrization.
Our starting point was the force-field parametrization (Fit\_A) derived in Piomponi \emph{et al} \cite{piomponi2022molecular},
which was able to reproduce optical melting experiments measuring the impact of the methylation on RNA-only structures.
Notably, for YTH these parameters led to a high overestimation of the methylation effect on protein-RNA binding.
We first evaluated possible sampling issues related to the hydration of the binding pocket \cite{krepl2021recognition}. Alchemical simulations of water annihilation and alchemical simulations of A methylation coupled with metadynamics were used to quantify this effect, which resulted to be insufficient to explain the discrepancy with experiment.
We then evaluated the effect of the force-field parametrization, computing $\Delta \Delta G_{bind}$ for alternative m$^6$A force-field, including the Aduri force-field and the parametrization used by Krepl \emph{et al}.
The results showed that the Krepl and Aduri force-field destabilized the YTH-RNA complex compared to Fit\_A. We attributed this to Fit\_A having more polar H61, N3 and N1 atoms, which formed hydrogen bonds in the aromatic cage. Although  Aduri  force-field can reproduce experimental $\Delta \Delta G_{bind}$, none of the so far considered m$^6A$ force-field could simoultaneosuly replicate isomer populations, duplex denaturation experiments, and calorimetry experiments on the YTH-RNA complex simultaneously.
To address these issues, we extended the fitting procedure used in our previous work \cite{piomponi2022molecular}, using an expanded experimental dataset, that includes the YTH-RNA $\Delta \Delta G_{bind}$. 
The newly fitted parameters (fit5\_AC) do not effectively reproduce experimental $\Delta \Delta G_{bind}$ as well as  other alternative parametrizations, but offer a balanced compromise between matching $\Delta \Delta G_{bind}$ and denaturation experiments. This is due to the adjustments made to the polarity of N1, H61, and N3 atoms, which impact hydrogen bonding and stability in different systems.

An important ingredient in our work is the employment of metadynamics coupled with alchemical simulations.
This combination has been already proposed in a number of different flavors by different authors \cite{hsu2023alchemical,khuttan2024make}.
In our case, these calculations enabled us to rule out a possibly important contribution of hydration
to the free energy of binding and led us to the crucial finding that experimental data could only be reproduced by properly tuning the m$^6$A force-field
parameters. Adjusting force-field to experiment is becoming more and more common
\cite{cesari2019fitting,frohlking2020toward,kummerer2023fitting,bolhuis2023optimizing,gilardoni2024boosting}, but the idea of tuning charges has been tested only recently \cite{piomponi2022molecular}.
Importantly, binding free energies are highly sensitive to partial charges. This has a two-fold implication: on the one hand,
a small deviation in the employed parameters can lead to gross inaccuracies; on the other hand,
a limited number of experimental observations could provide very strict bounds on these parameters.
Our results also reinforce the fact that experimental data used in training force-field parameters
should include as many interactions as possible. In our specific case, a dataset probing only
interactions formed by one edge of the nucleobase was not able to identify problems with charges present for the other edges. 
Our findings confirm it is absolutely crucial to combine multiple and structurally diverse datasets when training
force-field on experimental data.

\section{Data availability}

Jupyter notebooks used for molecular dynamics simulations and analysis can be found at \url{https://github.com/bussilab/m6a-charge-fitting}.
Input files and trajectory data are available at \url{https://doi.org/10.5281/zenodo.11002098}.

\section{Funding}
V. P. was supported by the European Union – NextGenerationEU within the project PNRR (PRP@CERIC) IR0000028 - Mission 4 Component 2 Investment 3.1 Action 3.1.1. \\
J. S. and M.K. acknowledge support of Czech Science Foundation (grant number 23-05639S).

\bibliographystyle{unsrt}
\bibliography{main}%

\end{document}